\pdfoutput=1

\documentclass[11pt,twoside,a4paper,cmspaper,final,collab]{cms-tdr}
\def\svnVersion{434904P}\def\cmsCernNoTag{CERN-EP-2017-177}\def\cmsCernDate{\today}\def\cmsMessage{Published in Physics Letters B as \href{http://dx.doi.org/10.1016/j.physletb.2017.10.020}{\doi{10.1016/j.physletb.2017.10.020.}}}

\begin{document}\cmsNoteHeader{SMP-17-006}

\hyphenation{had-ron-i-za-tion}
\hyphenation{cal-or-i-me-ter}
\hyphenation{de-vices}
\RCS$HeadURL: svn+ssh://svn.cern.ch/reps/tdr2/papers/SMP-17-006/trunk/SMP-17-006.tex $
\RCS$Id: SMP-17-006.tex 426469 2017-09-21 10:36:25Z ppigard $

\newlength\cmsFigWidth
\ifthenelse{\boolean{cms@external}}{\setlength\cmsFigWidth{0.45\textwidth}}{\setlength\cmsFigWidth{0.55\textwidth}}
\ifthenelse{\boolean{cms@external}}{\providecommand{\cmsLeft}{top\xspace}}{\providecommand{\cmsLeft}{left\xspace}}
\ifthenelse{\boolean{cms@external}}{\providecommand{\cmsRight}{bottom\xspace}}{\providecommand{\cmsRight}{right\xspace}}
\newcommand{\x}{\ensuremath{\phantom{0}}}
\newcommand{\y}{\ensuremath{\phantom{.}}}

\cmsNoteHeader{SMP-17-006}

\title{Measurement of vector boson scattering and constraints on anomalous quartic couplings from events with four leptons and two jets in proton--proton collisions at $\sqrt{s}=13\TeV$}

\date{\today}

\newcommand{\MG}{\textsc{MG5\_aMC}\xspace} 
\newcommand{\ZZjj}{\ensuremath{\PZ\PZ\mathrm{jj}}\xspace} 

\abstract{A measurement of vector boson scattering and constraints on anomalous quartic gauge couplings from events with two Z bosons and two jets are presented. The analysis is based on a data sample of proton--proton collisions at $\sqrt{s}=13\TeV$ collected with the CMS detector and corresponding to an integrated luminosity of 35.9\fbinv. The search is performed in the fully leptonic final state $\PZ\PZ\to\ell\ell\ell^{\prime}\!\ell^{\prime}\!$, where $\ell,\, \ell^{\prime}\! = \Pe$ or $\mu$. The electroweak production of two Z bosons in association with two jets is measured with an observed (expected) significance of 2.7 (1.6) standard deviations. A fiducial cross section for the electroweak production is measured to be ${\sigma_{\textrm{EW}}(\Pp\Pp\rightarrow\PZ\PZ\textrm{jj}\rightarrow\ell\ell\ell^{\prime}\!\ell^{\prime}\textrm{jj}) = 0.40^{+0.21}_{-0.16}\stat~^{+0.13}_{-0.09}\syst\unit{fb}}$, which is consistent with the standard model prediction. Limits on anomalous quartic gauge couplings are determined in terms of the effective field theory operators T0, T1, T2, T8, and T9. This is the first measurement of vector boson scattering in the $\PZ\PZ$ channel at the LHC.}

\hypersetup{%
pdfauthor={CMS Collaboration},%
pdftitle={Measurement of vector boson scattering and constraints on anomalous quartic couplings from events with four leptons and two jets in proton-proton collisions at sqrt(s) = 13 TeV},%
pdfsubject={CMS},%
pdfkeywords={CMS, physics, SM, ZZ, VBS, aQGC}}

\maketitle 

\section{Introduction}
\label{sec:introduction}

Weak vector boson scattering (VBS) plays a central role in the standard model (SM) and is a key process to probe the non-Abelian gauge structure of the electroweak (EW) interaction. In the absence of any other contributions, the scattering amplitude of longitudinally polarized vector bosons would violate unitarity at center-of-mass energies for the scattering process of order 1\TeV~\cite{Lee:1977yc,Lee:1977eg}.
The discovery of a scalar boson at the CERN LHC~\cite{Aad:2012tfa,Chatrchyan:2012xdj} with gauge couplings compatible with those predicted for the SM Higgs boson~\cite{Khachatryan:2016vau} provides evidence that contributions from the exchange of this boson may be responsible for preserving unitarity at high energies, as predicted in the SM. 

Unitarity restoration for longitudinal boson scattering in the SM relies on the interference of the VBS amplitudes and amplitudes that involve the Higgs boson. Any deviation in the SM coupling of the Higgs boson to the gauge bosons breaks this delicate cancellation, thus permitting a test of the EW symmetry breaking mechanism (EWSB) of the SM. The study of differential cross sections for VBS processes at large diboson invariant masses provides a model-independent test of the Higgs boson couplings to vector bosons, complementing direct measurements of Higgs boson production and decay rates. Many models of physics beyond the SM alter the couplings of vector bosons, and the effects can be parametrized in an effective field theory approach~\cite{Degrande:2012wf}. The VBS topology increases the sensitivity  to the contribution of the quartic interactions, allowing tests for the presence of anomalous quartic gauge couplings (aQGCs)~\cite{Eboli:2006wa}.

\begin{figure*}[t]
\begin{center}
\includegraphics[width=0.6\textwidth]{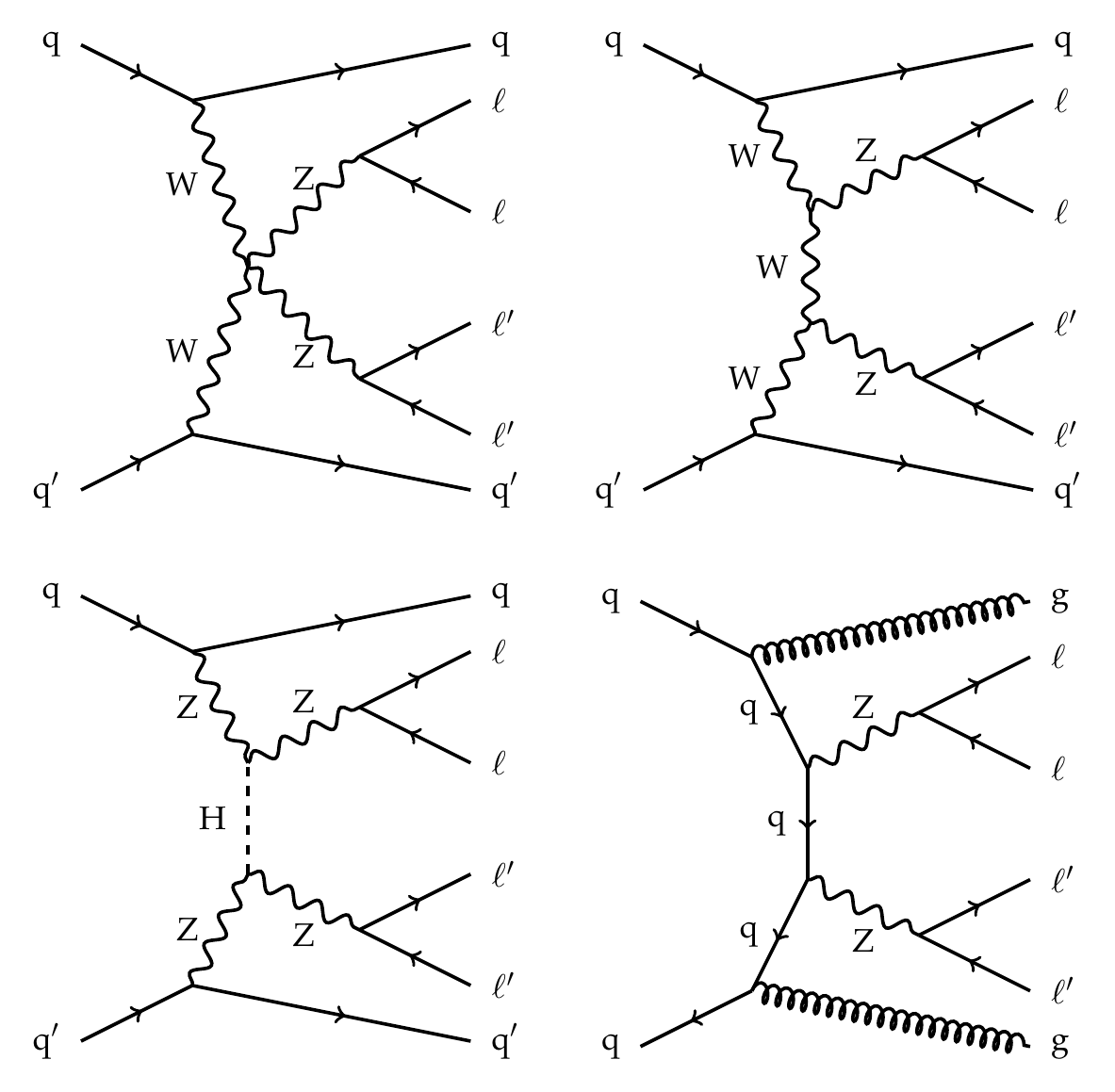}
\caption{Representative Feynman diagrams for the EW- (top row and bottom left) and QCD-induced production (bottom right) of the $\PZ\PZ \textrm{jj}\rightarrow\ell\ell\ell^{\prime}\!\ell^{\prime} \textrm{jj}$ ($\ell,\, \ell^{\prime}\!=\Pe$ or $\mu$) final state. The scattering of massive gauge bosons as depicted in the top row is unitarized by the interference with amplitudes that feature the Higgs boson (bottom left).} 
\label{fig:feynman}
\end{center}
\end{figure*}

At the LHC, VBS is initiated by quarks $\PQq$ from the colliding protons; both quarks radiate vector bosons ($\textrm{V}=\PW$, $\PZ$) which then interact. Because of the relatively small transverse momentum \pt carried by the gauge bosons and the absence of any color exchange at leading order (LO), VBS is characterized by the presence of two forward jets j in addition to the outgoing gauge bosons ($\PQq\PQq\rightarrow \textrm{VVjj}$) and little hadronic activity between the two jets~\cite{rapidity_gap_1, rapidity_gap_2}. The hard interaction in VBS only involves the EW interaction. Figure \ref{fig:feynman} shows some of the Feynman diagrams that contribute to the EW production of the $\textrm{VVjj}$ signature, involving quartic (top left) and trilinear vertices (top right), as well as diagrams involving the Higgs boson (bottom left). The $\PQq\PQq\rightarrow \textrm{VVjj}$ process can also be mediated through the strong interaction (bottom right in Fig.~\ref{fig:feynman}), which leads to the same final state as the VBS signal, resulting in an irreducible background.

Both the ATLAS and CMS Collaborations performed searches for VBS using proton--proton collisions at $\sqrt{s}=8\TeV$, notably in the same-sign WW channel~\cite{ATLAS_ssWW, CMS_ssWW,  Aaboud:2016ffv}. The ATLAS Collaboration also reported limits on a fiducial cross section for VBS in the WZ channel \cite{Aad:2016ett}. The ZZ channel remained unprobed. Limits on aQGCs are reported in Refs.~\cite{ATLAS_ssWW, CMS_ssWW,  Aaboud:2016ffv,Aad:2016ett, Khachatryan:2016mud, Aaboud:2016uuk, Khachatryan:2017jub,  Khachatryan:2016vif,Aaboud:2017pds}. 

This paper presents the first experimental investigation of VBS in the ZZ channel and exploits the fully leptonic final state, where both $\PZ$ bosons decay into electrons or muons, $\PZ\PZ\rightarrow\ell\ell\ell^{\prime}\!\ell^{\prime}\!$ ($\ell,\, \ell^{\prime}\!=\Pe$ or $\mu$). Despite a low cross section, a small $\PZ\rightarrow\ell\ell$ branching fraction, and a large irreducible QCD background, this channel provides a favorable laboratory to study EWSB because all final-state particles are reconstructed.  The clean leptonic final state results in a small reducible background, where one or more of the reconstructed lepton candidates originate from the misidentification of jet fragments. This channel also provides a precise knowledge of the scattering energy. Furthermore, the spin correlations of the reconstructed fermions permit the extraction of the longitudinal contribution to VBS.

The search for the EW production of the $\ell\ell\ell^{\prime}\!\ell^{\prime} \textrm{jj}$ final state is carried out using pp collisions at $\sqrt{s}=13\TeV$ recorded with the CMS detector at the LHC. The data set corresponds to an integrated luminosity of 35.9\fbinv collected in 2016. A multivariate discriminant, which combines observables sensitive to the kinematics of the VBS process to separate the EW- from the QCD-induced production, is used to extract the signal significance and to measure the cross section for the EW production in a fiducial volume. Finally, the selected $\ell\ell\ell^{\prime}\!\ell^{\prime} \textrm{jj}$ events are used to constrain aQGCs described by the operators T0, T1, and T2 as well as the neutral-current operators T8 and T9~\cite{Eboli:2006wa}. 

\section{The CMS detector}
\label{sec:cmsandselection}

The central feature of the CMS apparatus is a superconducting solenoid of 6\unit{m} internal diameter, providing a magnetic field of 3.8\unit{T}. Within the solenoid volume are silicon pixel and strip tracking detectors, a lead tungstate crystal electromagnetic calorimeter (ECAL), and a brass and scintillator hadron calorimeter (HCAL), each composed of a barrel and two endcap sections. Forward calorimeters extend the pseudorapidity $\eta$ coverage provided by the barrel and endcap detectors up to $\abs{\eta} < 5$. Muons are measured in gas-ionization detectors embedded in the steel flux-return yoke outside the solenoid.

The silicon tracker measures charged particles within the pseudorapidity range $\abs{\eta}< 2.5$. It consists of 1440 silicon pixel and 15\,148 silicon strip detector modules. For nonisolated particles with $1 < \pt < 10\GeV$ and $\abs{\eta} < 1.4$, the track resolutions are typically 1.5\% in \pt and 25--90 (45--150)\mum in the transverse (longitudinal) impact parameter~\cite{TRK-11-001}.

Electrons are measured in the pseudorapidity range $|\eta|~{<}~2.5$ using both the tracking system and the ECAL. 
The momentum resolution for electrons with $\pt \approx 45\GeV$ from $\PZ \rightarrow \Pe^+ \Pe^-$ decays ranges from 1.7\% for nonshowering electrons in the barrel region ($|\eta|<1.479$) to 4.5\% for showering electrons in the endcaps~\cite{Khachatryan:2015hwa}.

Muons are measured in the pseudorapidity range $|\eta|~{<}~2.4$ using the silicon tracker and muon systems. The muon detectors are constructed using three different technologies: drift tubes for $|\eta| < 1.2$, cathode strip chambers for $0.9 < |\eta| < 2.4$, and resistive plate chambers for $|\eta| < 1.6$. In the intermediate \pt range of $20<\pt<100\GeV$, matching muons to tracks measured in the silicon tracker results in a relative \pt resolution of 1.3--2.0\% in the barrel ($|\eta|<1.2$), and better than 6\% in the endcaps. The \pt resolution in the barrel is better than 10\% for muons with \pt up to 1\TeV~\cite{Chatrchyan:2012xi}.

In the region $\abs{ \eta } < 1.74$, the HCAL cells have widths of 0.087 in pseudorapidity and 0.087 in azimuth ($\phi$). 
In the $\eta$-$\phi$ plane, and for $\abs{\eta} < 1.48$, the HCAL cells map on to $5{\times}5$ arrays of ECAL crystals to form calorimeter towers projecting radially outwards from the nominal interaction point. For $\abs{\eta} > 1.74$, the size of the towers increases progressively to a maximum of 0.174 in $\Delta \eta$ and $\Delta \phi$. When combining information from the entire detector, the jet energy resolution amounts typically to 15\% at 10\GeV, 8\% at 100\GeV, and 4\% at 1\TeV, to be compared to about 40\%, 12\%, and 5\% obtained when the ECAL and HCAL calorimeters alone are used.

The first level of the CMS trigger system, composed of custom hardware processors, uses information from the calorimeters and muon detectors to select events of interest in a fixed time interval of 3.2\mus. The high-level trigger processor farm further decreases the event rate from around 100\unit{kHz} to less than 1\unit{kHz}, before data storage~\cite{Khachatryan:2016bia}.

A more detailed description of the CMS detector, together with a definition of the coordinate system used and the relevant kinematic variables, can be found in Ref.~\cite{Chatrchyan:2008zzk}.

\section{Signal and background simulation}
\label{sec:mc}

Several Monte Carlo event generators are used to simulate the signal and background contributions.
The simulated samples are employed to optimize the event selection, to develop the multivariate discriminator, and to
estimate the irreducible background yields.

The EW production of Z boson pairs and two final-state quarks, where the Z bosons decay leptonically, is simulated at LO using 
\MADGRAPH{}5\_a\MCATNLO~v2.3.3 (abbreviated as \MG in the following)~\cite{MGatNLO}. The sample includes triboson processes, where the Z boson pair is accompanied by a third vector boson that decays into jets, as well as 
diagrams involving the quartic coupling vertex. The predictions from this sample are cross-checked with those obtained from the LO generator \textsc{Phantom}\ v1.2.8~\cite{Phantom}, and excellent agreement in the yields and the multivariate distribution exploited for the signal extraction is found.

The event samples of the QCD-induced production
of two $\PZ$ bosons are simulated with zero, one, and two outgoing partons at Born level at next-to-leading order (NLO) with 
\MG. The different jet multiplicities are merged using the FxFx scheme~\cite{Frederix:2012ps} with a merging scale of 30\GeV, and leptonic $\PZ$ boson decays are simulated using \textsc{MadSpin}~\cite{Artoisenet:2012st}.
The interference between the EW and QCD diagrams is evaluated using dedicated samples produced with \MG at LO. It is found to contribute less than 1\% to the total yield and is therefore neglected.
The loop-induced production of two \PZ bosons, referred to as $\Pg\Pg \to\PZ\PZ$, is simulated at LO with \MCFM~v.7.0.1~\cite{MCFM}. A dedicated \MG simulation of the loop-induced $\Pg\Pg\to\ZZjj$  process is used to check the modeling of the \ZZjj phase space in the \MCFM sample, and good agreement is found.

Samples for $\ttbar\PZ$ and $\PW\PW\PZ$ production, background processes that contain four prompt, isolated leptons and additional jets in the
final state, are simulated with \MG at NLO.  

The simulation of the aQGC processes is performed at LO using \MG and employs matrix element reweighting to obtain a finely spaced grid in each of the five anomalous couplings probed by the analysis.

The \PYTHIA~v8.212~\cite{Sjostrand:2006za,Sjostrand:2015} package
is used for parton showering, hadronization, and
the underlying event simulation, with parameters set by the CUETP8M1
tune~\cite{Khachatryan:2015pea}. The NNPDF3.0~\cite{NNPDF2015} set of
parton distribution functions (PDFs) is used, and the PDFs are calculated to the same
order in QCD as the hard process. All simulated samples are normalized 
to the cross sections obtained from the respective event generator.

The detector response is simulated using a detailed
description of the CMS detector implemented in the \GEANTfour
package~\cite{GEANT, Geant2}. The simulated events are  reconstructed
using the same algorithms as used for the data. 
The simulated samples include additional interactions in the same and neighboring bunch crossings,
referred to as pileup.
Simulated events are weighted so that the pileup distribution reproduces that observed in 
the data, which has an average of about 23 interactions per bunch
crossing.

\section{Event selection}
\label{sec:eventselection}

The final state should consist of at least two pairs of oppositely charged isolated leptons and at least two hadronic jets. The ZZ selection is similar to that used in the CMS inclusive ZZ cross section measurement~\cite{ZZXS13TeV2015Paper}. 

The primary triggers require the presence of a
pair of loosely isolated leptons.
The highest \pt electron~(muon) must have $\pt^\ell > 23~(17)\GeV$, and the
next-to-highest \pt lepton must have
$\pt^\ell > 12~(8)\GeV$. The dilepton triggers require that the
tracks associated with the leptons originate from within
2\mm of each other along the beam axis. Triggers
requiring a triplet of low-\pt leptons
with no isolation criterion, as well as isolated single-electron and single-muon triggers with minimal \pt thresholds of 27 and 22\GeV, respectively, help to recover efficiency.
The overall trigger efficiency for events that satisfy the ZZ selection described below is greater than 98\%.

Events are reconstructed using a particle-flow algorithm~\cite{CMS-PRF-14-001} that reconstructs and identifies each individual particle with an optimized combination of all subdetector information. The missing transverse momentum vector \ptvecmiss is defined as the projection onto the plane perpendicular to the beam axis of the negative vector sum of the momenta of all reconstructed particle-flow objects in an event. Its magnitude is referred to as \ptmiss.

The reconstructed vertex with the largest value of summed physics-object $\pt^2$ is taken to be the primary $\Pp\Pp$ interaction vertex. The physics objects are the objects returned by a jet finding algorithm~\cite{Cacciari:2008gp,Cacciari:2011ma} applied to all charged tracks associated with the vertex, plus the corresponding associated \ptmiss. Leptons and jets are required to originate from the primary vertex. 

Electrons are identified using a multivariate classifier, which includes observables sensitive to 
bremsstrahlung along the electron trajectory, the geometrical and energy-momentum compatibility between the 
electron track and the associated energy cluster in the ECAL, the shape of the electromagnetic shower, 
and variables that discriminate against electrons originating from photon conversions~\cite{Khachatryan:2015hwa}. 

Muons are reconstructed by combining 
information from the silicon tracker and the muon system~\cite{Chatrchyan:2012xi}.
The matching between the inner and outer tracks proceeds either outside-in, starting from a track in the muon system, 
or inside-out, starting from a track in the silicon tracker. 
The muons are selected from the reconstructed muon track candidates by applying minimal requirements on the track in both 
the muon system and silicon tracker, and taking into account compatibility with small energy deposits in the calorimeters.

In order to suppress electrons from photon conversions and muons originating from in-flight decays of hadrons, we require the three-dimensional impact parameter of each lepton track, computed with respect to the primary vertex position, to be less than four times the uncertainty on the impact parameter.

Leptons are required to be isolated from other particles in the event. The
relative isolation is defined as
\begin{equation}
R_\text{iso} = \bigg[ \sum_{\substack{\text{charged} \\ \text{hadrons}}} \!\! \pt \, + \,
\max\big(0, \sum_{\substack{\text{neutral} \\ \text{hadrons}}} \!\! \pt
\, + \, \sum_{\text{photons}} \!\! \pt \, - \, \pt^\mathrm{PU}
\big)\bigg] \bigg/ \pt^{\ell},
\label{eq:iso}
\end{equation}
where the sums run over the charged and neutral hadrons and photons, in a
cone defined by
$\Delta R \equiv \sqrt{\smash[b]{\left(\Delta\eta\right)^2 + \left(\Delta\phi\right)^2}} = 0.3$
around the lepton trajectory.
To minimize the contribution of charged particles from pileup to the isolation calculation,
charged hadrons are included only if they originate from the
primary vertex. The contribution of
neutral particles from pileup is $\pt^\mathrm{PU}$. For electrons, $\pt^\mathrm{PU}$
is evaluated with the jet area method described in Ref.~\cite{Cacciari:2007fd}.
For muons, $\pt^\mathrm{PU}$ is taken
to be half the \pt sum of all charged particles in the cone originating
from pileup vertices. The factor of
one-half accounts for the expected ratio of charged to neutral particle energy
in hadronic interactions. Leptons with $R_\text{iso} < 0.35$ are considered isolated.

The efficiency of the lepton reconstruction and selection is 
measured in bins of $\pt^\ell$ 
and $\eta^\ell$ using the tag-and-probe technique. The measured efficiencies are used to correct the simulation.
The lepton momentum scales are calibrated in bins of $\pt^\ell$ and $\eta^\ell$ 
using the decay products of known dilepton resonances. The electron momentum 
scale for data is corrected with a $\PZ\rightarrow \Pe^+\Pe^-$ sample by 
matching the peak of the reconstructed dielectron mass spectrum to the known 
value of $m_{\PZ}$. A Gaussian smearing of the electron energies 
in the simulation is also applied to match the $\PZ\rightarrow \Pe^+\Pe^-$ mass resolution 
in data. Muon momenta are calibrated using a Kalman filter approach~\cite{Fruhwirth:1987fm}, using \JPsi meson
and \PZ boson decays. 

An algorithm is used to identify final-state radiation (FSR) from the
leptons~\cite{Chatrchyan:2013mxa}. 
A photon with ${\pt>2\GeV}$ and within a cone of $\Delta R=0.5$ around the lepton momentum direction is selected if it satisfies quality requirements. 
The FSR photons identified by the algorithm are excluded
from the lepton isolation computation.

Jets are reconstructed from particle-flow candidates using the anti-\kt clustering algorithm~\cite{Cacciari:2008gp}, as implemented in the \textsc{FASTJET} package~\cite{Cacciari:2011ma}, with a distance parameter of 0.4. In order to assure a good reconstruction efficiency and to reduce the instrumental background as well as the contamination from pileup, loose identification criteria based on the multiplicities and energy fractions carried by charged and neutral hadrons are imposed on jets~\cite{CMS-PAS-JME-16-003}. Only jets with $|\eta| < 4.7$ are considered.

Jet energy corrections are extracted from data and simulated events to account for the effects of pileup, uniformity of the detector response, and residual differences between the jet energy scale in the data and in the simulation.
The jet energy scale calibration~\cite{1748-0221-6-11-P11002, Khachatryan:2016kdb, JESCalib} relies on corrections parameterized in terms of the uncorrected \pt and $\eta$ of the jet, and is applied as a multiplicative factor, scaling the four-momentum vector of each jet. In order to ensure that jets are well measured and to reduce the pileup contamination, all jets must have a corrected \pt larger than 30\GeV.

A signal event must contain at least two
$\PZ$ candidates, each formed from pairs of isolated electrons or muons of opposite charges.
Only reconstructed electrons (muons) with a $\pt > 7~(5) \GeV$ are considered.
Among the four leptons, the highest \pt lepton must have $\pt > 20\GeV$, and
the second-highest \pt lepton must have $\pt > 12~(10) \GeV$ if it is an electron (muon).
All leptons are required to be separated by
$\Delta R \left(\ell_1, \ell_2 \right) > 0.02$, and electrons are required to be separated from muons by
$\Delta R \left(\Pe, \mu \right) > 0.05$.

Within each event, all
permutations of leptons giving a valid pair of $\PZ$
candidates are considered.
For each ZZ candidate, the
lepton pair with the invariant mass closest to the nominal Z boson mass is denoted $\PZ_1$ 
and is required to have a mass greater than 40\GeV. The other dilepton candidate is denoted $\PZ_2$. Both $m_{\PZ_1}$ and $m_{\PZ_2}$ are required to be less than 120\GeV.
All pairs of oppositely charged leptons, regardless of flavor, in the ZZ candidate are required to
satisfy ${m_{\ell \ell^{\prime}\!} > 4\GeV}$ to suppress backgrounds from hadron decays.

If multiple $\PZ\PZ$
candidates in an event pass this selection,
the candidate with $m_{\PZ_1}$ closest to
the nominal $\PZ$ boson mass is chosen. In the rare case (0.3\%)
of further ambiguity, which may arise in events with
more than four leptons, the $\PZ_2$ candidate
that maximizes the scalar $\pt$ sum of the four leptons is chosen. 
Finally, the $\PZ_1$ and $\PZ_2$ candidates must have masses between $60$ and
$120\GeV$. This selection is referred to as the $\PZ\PZ$ selection.

The search for the EW production of two $\PZ$ bosons is performed on a 
subset of events that pass the $\PZ\PZ$ selection, namely those that feature at least two jets. The jets are required to be separated from the leptons of the $\PZ\PZ$ candidate by $\Delta R=0.4$. The two highest \pt jets 
are referred to as the tagging jets and their invariant mass is 
required to be larger than 100\GeV. This selection is referred to as the \ZZjj selection.

\section{Background estimation}
\label{sec:bckgestimation}

The dominant background is the QCD-induced production of two \PZ bosons in association with jets, as shown in the bottom right diagram of Fig.~\ref{fig:feynman}. The yield and shape of the multivariate discriminant of this irreducible background are taken from simulation, but ultimately constrained by the data in the fit that extracts the EW signal, as described in Section~\ref{sec:VBS}. Other irreducible backgrounds arise from processes that produce four genuine 
high-\pt isolated leptons, $\Pp\Pp\to\ttbar\PZ\textrm{\,+\,jets}$ and $\Pp\Pp\to\PW\PW\PZ\textrm{\,+\,jets}$.
These small contributions feature kinematic distributions similar to that of the dominant background and are estimated using simulation.

Reducible backgrounds arise from processes in which heavy-flavor jets produce 
secondary leptons or from processes in which jets are misidentified as leptons. 
The lepton identification and isolation requirements significantly suppress this
background, which is very small compared to the signal after the selection. 

The reducible background, referred to as $\PZ\textrm{\,+\,X}$, is predominately composed of $\PZ\textrm{\,+\,jets}$ events, with minor contributions from $\ttbar\textrm{\,+\,jets}$ and $\PW\PZ\textrm{\,+\,jets}$ processes. 
This reducible contribution is estimated from data by inverting the lepton selection criteria and weighting events in control regions using a lepton misidentification rate which is also determined from data.  Two control regions serve to estimate the reducible background from events with one or two misidentified leptons, respectively.

Events in the control region with one (two) misidentified lepton(s) satisfy the \ZZjj selection, with the exception that one of the \PZ boson candidates is constructed from one (two) lepton(s) that fail the identification or isolation criteria.
The lepton misidentification rate is measured by selecting events that feature one \PZ boson candidate and a third reconstructed lepton. 
The fraction of events for which the third lepton satisfies the identification and isolation criteria is taken as the lepton misidentification rate.
The procedure is identical to that used in 
Ref.~\cite{ZZXS13TeV2015Paper} and is described in more detail in Ref.~\cite{Chatrchyan:2013mxa}.

\section{Systematic uncertainties}
\label{sec:syst}

Several sources of systematic uncertainty are considered and evaluated by varying each relevant parameter. The resulting
changes to the distribution of the multivariate discriminant, both in shape and yield, are taken into account. The impact of the variation for each source of uncertainty is summarized below.

Renormalization and factorization scale uncertainties are evaluated by varying both scales independently by factors of two and one-half, removing combinations where both variations differ by a factor of four, and amount to 10 (7)\% for the dominant QCD background (EW signal). The $\textrm{PDF\,+\,}\alpha_s$ variations are evaluated following the PDF4LHC prescription~\cite{PDFLHC}, and increase from 6\% at low values 
of the multivariate discriminant to 9\% in the signal-rich region. 
A 40\% uncertainty in the yield of the loop-induced \ZZjj background is assigned.
The impact of the jet energy scale 
uncertainty amounts to 20 (4)\% at low (high) values 
of the multivariate discriminant and the impact of the jet energy resolution uncertainty is 8\%~\cite{JESCalib, Khachatryan:2016kdb}. The uncertainties in the QCD background normalization and the jet energy scale are the dominant systematic uncertainties in the measurement. 
Higher order EW corrections in VBS processes are known to be negative and at the level of tens of percent~\cite{Biedermann:2016yds}, but such corrections have not been calculated for the final state considered in 
this paper, and therefore are not considered here. Nevertheless, the impact of such NLO EW corrections would be negligible in this analysis, which is limited by the large statistical uncertainty.
The uncertainty in the lepton reconstruction and selection efficiency is 6/4/2\% in the 4e/2e2$\mu$/4$\mu$ final states, respectively.
The uncertainty in the integrated luminosity is 2.5\%~\cite{CMS:2017sdi}. The systematic uncertainty in the trigger efficiencies is evaluated by taking the difference between the trigger efficiencies measured in data and in simulated events, and amounts to 2\%. A 40\% yield uncertainty in the reducible background estimate based on data samples takes into  account the limited number of events
in the control regions as well as the mismatch in the background composition in the control regions used to determine the lepton misidentification rates and the control regions used to estimate the yield in the signal region.

\section{Search for EW ZZjj production}
\label{sec:VBS}

\begin{figure*}[!htb]
\centering
\includegraphics[width=0.40\textwidth]{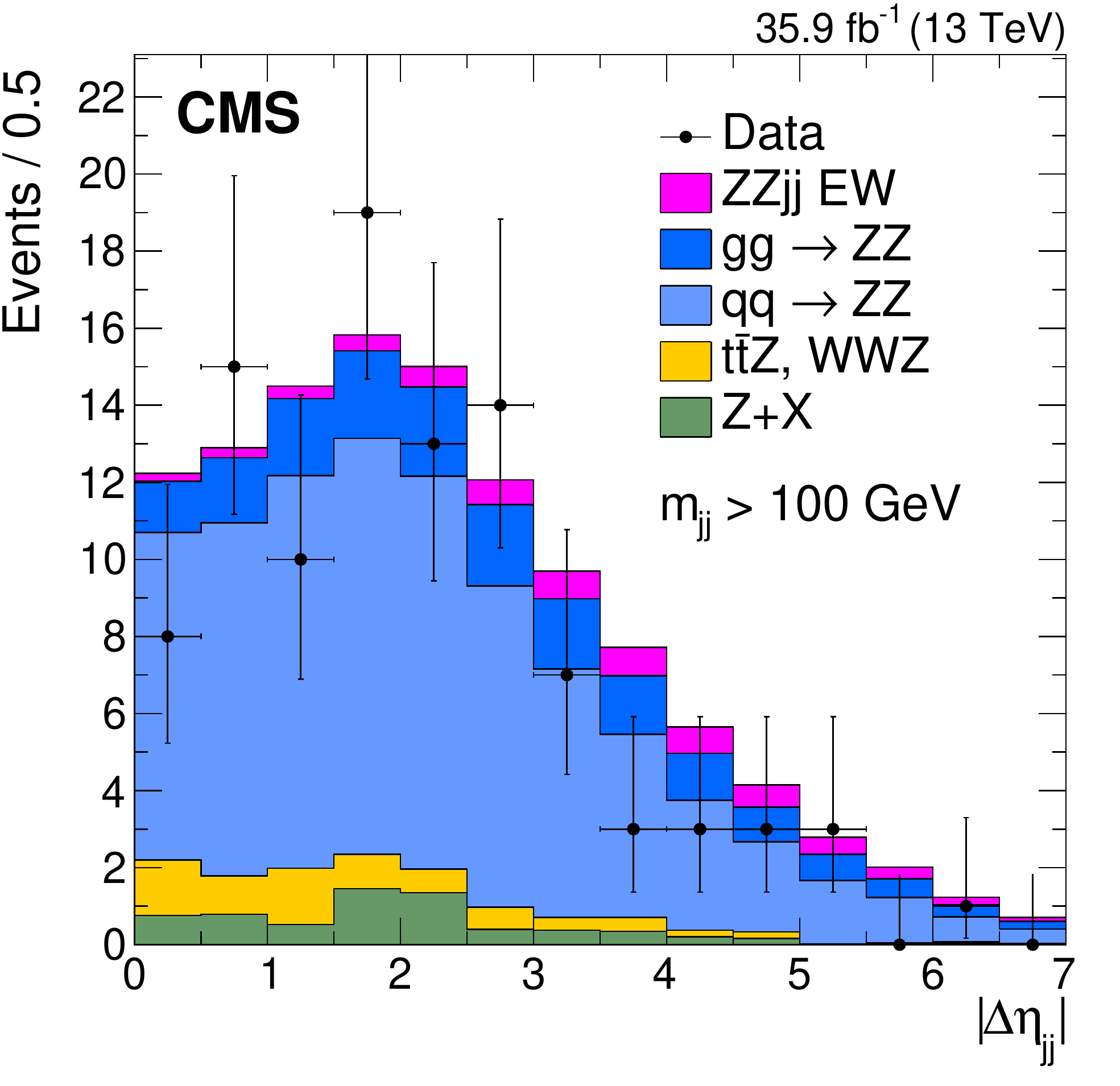}
\includegraphics[width=0.40\textwidth]{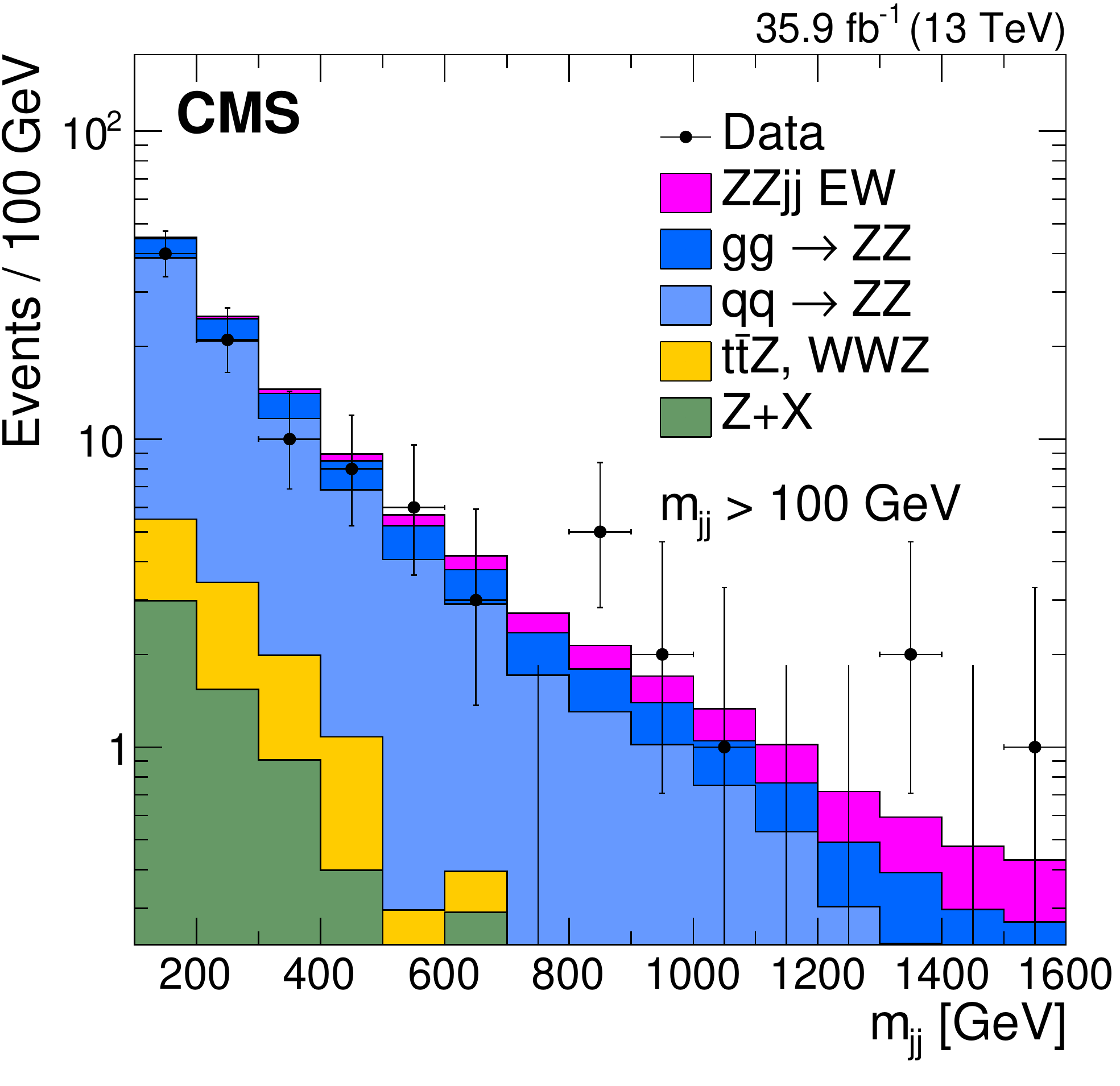}
\caption{Distribution of the dijet pseudorapidity separation (left) and dijet invariant mass (right) for events passing the \ZZjj selection, which requires $m_\textrm{jj} > 100\GeV$. Points represent the data, filled histograms the expected signal and background contributions. No data beyond $|\Delta\eta_\textrm{jj}|>7$ (left) and $m_\textrm{jj}>1600\GeV$ (right) is observed.}
\label{fig:simple_data_mc}
\end{figure*}

The expected signal purity in the \ZZjj selection is about 5\%, with 83\% of 
events coming from QCD-induced production. Additional kinematic 
selections are therefore necessary to enhance the contribution from EW production. 
Figure~\ref{fig:simple_data_mc} shows the absolute dijet pseudorapidity separation $|\Delta\eta_{\textrm{jj}}|$  
and the dijet invariant mass $m_\textrm{jj}$ for events passing 
the \ZZjj selection. Table~\ref{tab:res_yields} shows the expected
and observed number of events for the \ZZjj selection and illustrates the increase of the VBS signal purity obtained with an exemplary selection that requires ${m_\textrm{jj}>400\GeV}$ and ${|\Delta\eta_\textrm{jj}|>2.4}$.

\begin{table*}[!h]
\centering
\topcaption{Signal and background yields for the \ZZjj selection and for an illustrative VBS signal-enriched selection that requires $m_\textrm{jj}>400\GeV$ and $|\Delta\eta_\textrm{jj}|>2.4$ .}
\resizebox{\textwidth}{!}
{
\begin{tabular}{l c c c c c c c}
\hline 
Selection		&  $\ttbar\PZ$ and $\PW\PW\PZ$	& QCD \ZZjj 	& $\PZ\textrm{\,+\,X}$		& Total bkg. 	& EW \ZZjj	& Total expected & Data\\
\hline                                                                                                                                  
\ZZjj            & 
$   7.1 \pm   0.8$      & $  97\pm  14$ & $   6.6 \pm   2.5$ & $ 111 \pm  14$ & $   6.2 \pm   0.7$   & $ 117 \pm  14$   & $99$\\
VBS signal-enriched             & 
$   0.9 \pm   0.2$      & $  19 \pm 4\x $ & $   0.7 \pm   0.3$ & $  20 \pm   4$ & $   \x\y4 \pm   0.5$   & $  25 \pm   4$   & $19$\\
\hline
\end{tabular}
}
\label{tab:res_yields}
\end{table*}

The determination of the signal strength for the EW production, i.e., the ratio of the measured cross section to the SM expectation $\mu=\sigma/\sigma_{\textrm{SM}}$, employs a 
multivariate discriminant to optimally separate the signal and the QCD background. 
The scikit-learn framework \cite{scikit-learn} is used to train and optimize a 
boosted decision tree (BDT) on simulated events to exploit the kinematic differences
between the EW signal and the QCD background. Seven observables are used in the BDT, including $m_\textrm{jj}$, $|\Delta\eta_\textrm{jj}|$, $m_{{\PZ\PZ}}$, 
as well as the Zeppenfeld variables~\cite{rapidity_gap_1} $\eta^*_{\PZ_{i}}=\eta_{\PZ_{i}} - (\eta_{\textrm{jet 1}} + \eta_\textrm{jet 2})/2$ of the two \PZ bosons,
and the ratio between the \pt of the tagging jet system and the scalar \pt sum of the tagging jets.  
The BDT also exploits the event balance $R\pt^{\textrm{hard}}$, which is defined as the transverse component of the vector sum of the \PZ bosons and tagging jets momenta, normalized to the scalar \pt sum of the same objects \cite{Khachatryan:2014dea}. 

A total of 36 discriminating variables including observables sensitive to parton emissions between the tagging jets, the production and decay angles of the leptons, Z bosons, and tagging jets as well as quark-gluon tagging information are considered in the BDT training. Observables that do not improve the area under the signal-versus-background efficiency curve (AUC) are removed from the BDT. The observables sensitive to extra parton emissions provide little marginal AUC increase and are not retained because of the limited modelling accuracy in the simulation. 
The tunable hyper-parameters of the BDT training algorithm are optimized via a grid-search algorithm. Finally, the BDT performance is checked using a matrix element approach~\cite{Gao:2010qx,Bolognesi:2012mm,Anderson:2013afp} that provides a similar separation between the signal and background processes. 

\begin{figure*}[!htb]
\centering
\includegraphics[width=0.40\textwidth]{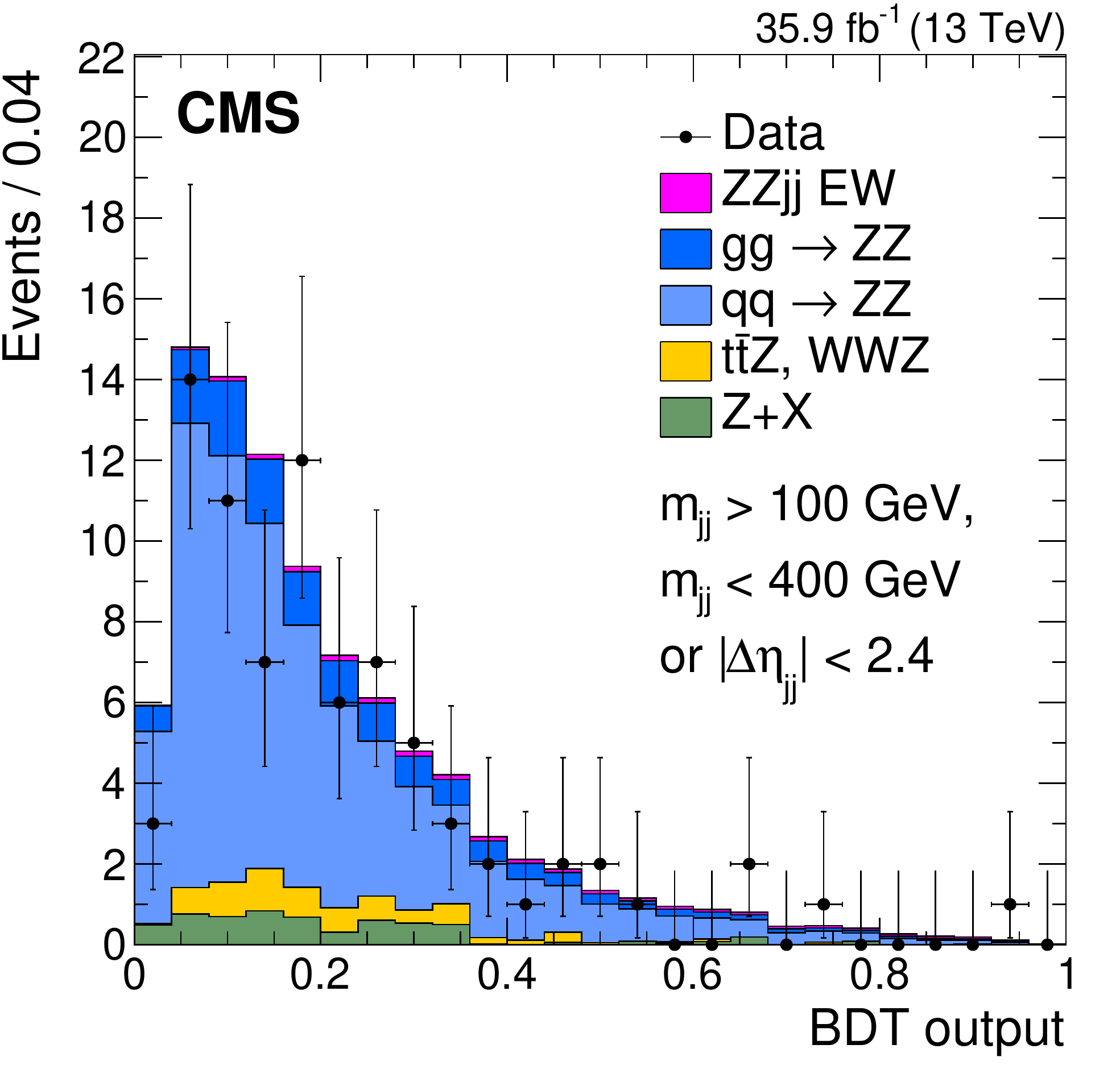}
\includegraphics[width=0.40\textwidth]{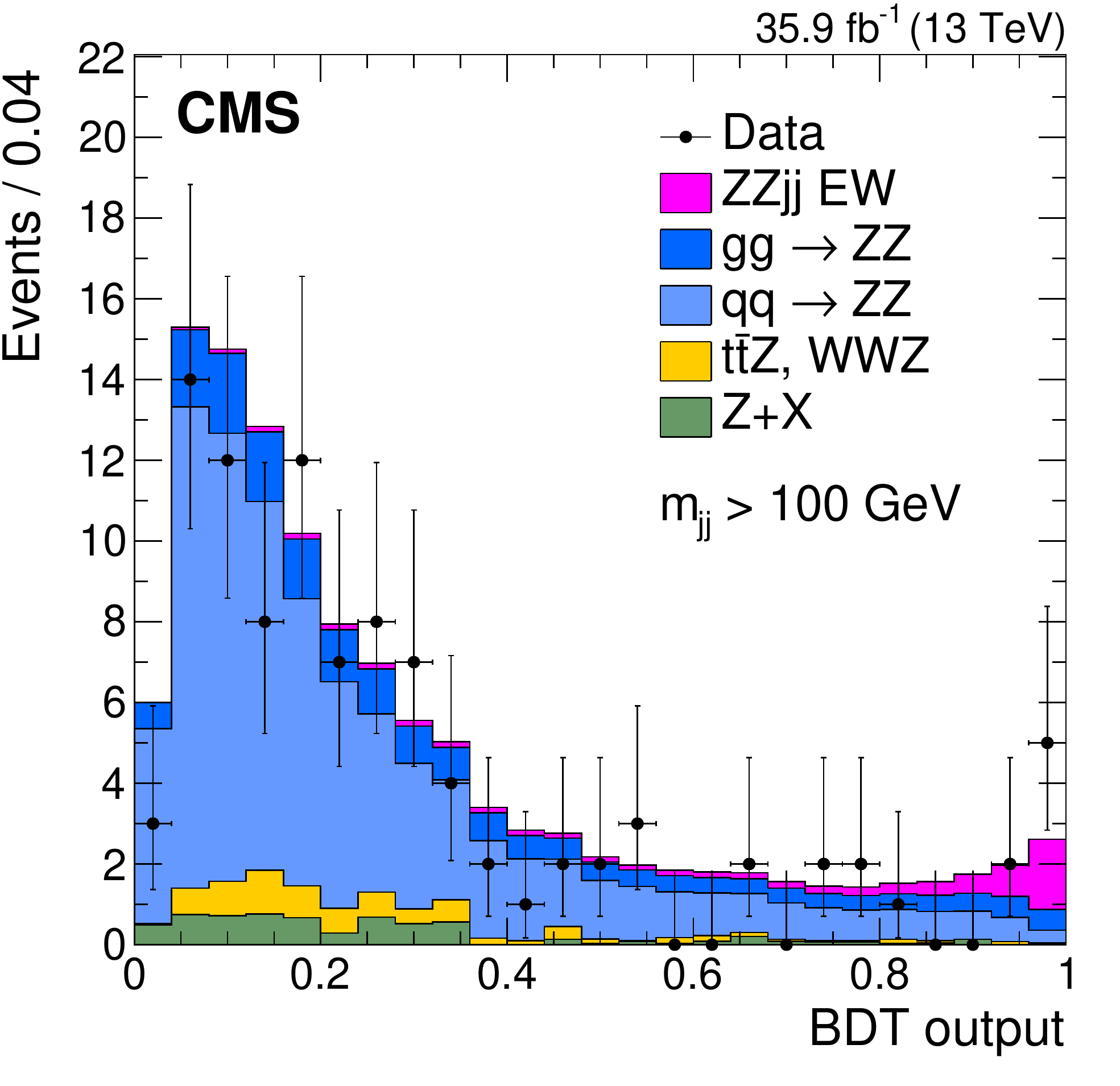}
\caption{Distribution of  the BDT output in the control region obtained by selecting \ZZjj events with $m_\textrm{jj}<400\GeV$ or $|\Delta\eta_\textrm{jj}|<2.4$ (left) 
and for the \ZZjj selection (right). Points represent the data, filled histograms the expected signal and background contributions.}
\label{fig:res_data_mc}
\end{figure*}

To validate the modeling of the backgrounds in the search, a QCD-enriched control region 
is defined by selecting events with $m_\textrm{jj}<400\GeV$ or $|\Delta\eta_\textrm{jj}|<2.4$. Good agreement is observed between the data and SM expectation in this control region, as shown in Fig.~\ref{fig:res_data_mc} (left). The classifier output distribution for all events in the \ZZjj selection including the high signal purity contribution at large BDT output values is shown in Fig.~\ref{fig:res_data_mc} (right).

\begin{figure*}[thb]
\centering
\includegraphics[width=\cmsFigWidth]{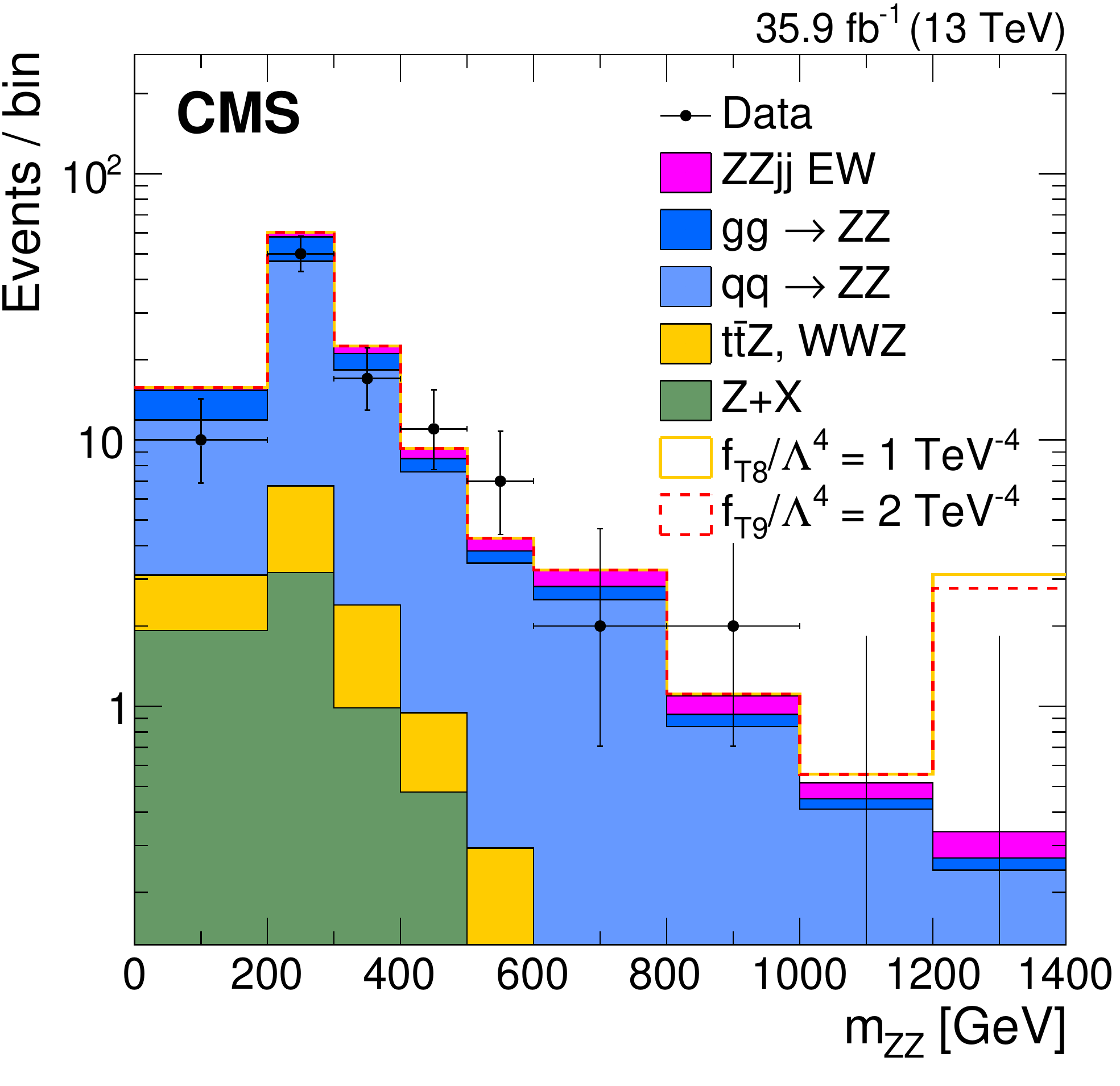}
\caption{The $m_{\PZ\PZ}$ distribution in the \ZZjj selection together with the SM 
prediction and two hypotheses for the aQGC coupling strengths. Points represent the data, filled histograms the expected signal and background contributions. The last bin includes all contributions with ${m_{\PZ\PZ}>1200\GeV}$.}
\label{fig:atgc}
\end{figure*}

The BDT distribution of the events in the \ZZjj selection is used to extract the significance of the EW 
signal via a maximum-likelihood fit. The expected distributions 
for the signal and the irreducible backgrounds are taken from the simulation while 
the reducible background is estimated from the data.
The shape and normalization of each distribution are allowed to vary in the fit within the respective uncertainties.
This approach constrains the yield of the QCD-induced production from the background-enriched region of the BDT distribution.

The systematic uncertainties are treated as nuisance parameters in the fit and profiled \cite{ATLAS:2011tau}. The post-fit values are then used to
extract the signal strength. The signal strength is measured to be  $\mu = 1.39^{+0.72}_{-0.57}\stat ~^{+0.46}_{-0.31}\syst = 1.39^{+0.86}_{-0.65}$ 
and the background-only hypothesis is excluded with a significance of $2.7$ standard deviations ($1.6$ standard deviations expected).

The measured signal strength is used to determine the fiducial cross section for the EW production. The fiducial volume is almost identical to the selections imposed at the reconstruction level, the only difference being the lepton thresholds of $\pt^\ell>5\GeV$ and $|\eta|^\ell<2.5$.
The generator-level lepton momenta are corrected by adding the momenta of generator-level photons within $\Delta R(\ell, \gamma) < 0.1$. The kinematic selection of the \PZ bosons and the final \ZZjj candidate proceeds as the reconstruction-level selection.
The observed signal strength corresponds to a fiducial cross section of  
${\sigma_{\textrm{EW}}(\Pp\Pp\rightarrow\textrm{ZZjj}\rightarrow\ell\ell\ell^{\prime}\!\ell^{\prime}\textrm{jj}) = 0.40^{+0.21}_{-0.16}\stat~^{+0.13}_{-0.09}\syst\unit{fb}}$, compatible with the SM prediction of ${0.29^{+0.02}_{-0.03}\unit{fb}}$.

\section{Limits on anomalous quartic gauge couplings}

The events in the \ZZjj selection are used to constrain aQGCs in the effective field theory approach. 
The \ZZjj channel is sensitive to the operators T0, T1, and T2, as well as the neutral current operators T8 and T9 \cite{Eboli:2006wa}.
The former operators are constructed from the SU$_L$(2) gauge fields, while the latter only involve the U$_Y$(1) fields. As a consequence, the T8 and T9 operators are experimentally accessible only  via final states involving the neutral gauge bosons.
The effect of a nonzero aQGC is to enhance the production cross section at large masses of the $\PZ\PZ$ system. Thus the $m_{\PZ\PZ}$ distribution is used to constrain the aQGC parameters $f_{\textrm{T}i}/\Lambda^4$. 
The increase of the yield exhibits a quadratic dependence on the anomalous coupling, and a parabolic function is fitted to the per-mass bin yields,
allowing for an interpolation between the discrete coupling parameters of the simulated signals.
The statistical analysis employs the same methodology used for 
the signal strength, including the profiling of the systematic uncertainties. 
The distributions of the background model, including the EW component, are normalized to their respective SM predictions.
The Wald Gaussian approximation and Wilks' theorem are used to derive 95\% confidence level (CL) limits on the aQGC parameters~\cite{CLS1, Junk:1999kv, CL_formula}. The measurement is statistically limited. 

\begin{table*}[!tb]
\centering
\topcaption{Expected and observed lower and upper 95\% CL limits on the couplings of the quartic operators T0, T1, and T2, as well as the neutral current operators T8 and T9. The unitarity bounds are also listed. All coupling parameter limits are in TeV$^{-4}$, while the unitarity bounds are in TeV.}
\begin{tabular}{l c c c c c}
\hline
Coupling			&  Exp. lower	& Exp. upper	& Obs.  lower	& Obs. upper	& Unitarity bound\\
\hline
$f_\textrm{T0}/\Lambda^4$       & $-0.53$	& $0.51$		& $-0.46$ 		& $0.44$         & $2.5$\\
$f_\textrm{T1}/\Lambda^4$       & $-0.72$	& $0.71$         	& $-0.61$		& $0.61$         & $2.3$\\
$f_\textrm{T2}/\Lambda^4$       & $-1.4\x$		& $1.4\x$          	& $-1.2\x$		& $1.2\x$          & $2.4$\\
$f_\textrm{T8}/\Lambda^4$       & $-0.99$	& $0.99$         	& $-0.84$		& $0.84$         & $2.8$\\
$f_\textrm{T9}/\Lambda^4$       & $-2.1\x$		& $2.1\x$          	& $-1.8\x$		& $1.8\x$          & $2.9$\\
\hline
\end{tabular}
\label{tab:aqgc_limits}
\end{table*}

Figure \ref{fig:atgc} 
shows the expected $m_{\PZ\PZ}$ distribution for the SM and
two aQGC scenarios. 
Table~\ref{tab:aqgc_limits} lists 
the individual lower and upper limits obtained by setting all other 
anomalous couplings to zero, as well as the unitarity bound. The unitarity bound 
is determined using the \textsc{VBFNLO} framework~\cite{VBFNLO} as the scattering energy $m_{\PZ\PZ}$ at which the aQGC coupling strength  set equal to the observed limit would result in a scattering amplitude that violates unitarity.
These are the most stringent limits to date on the aQGC parameters $f_{\textrm{T0,1,2}}/\Lambda^4$ and $f_{\textrm{T8,9}}/\Lambda^4$.

\ifthenelse{\boolean{cms@external}}{}{\clearpage}

\section{Summary}
\label{sec:summary}

A search was performed for vector boson scattering in the four-lepton and two-jet final
state using proton--proton collisions at 13\TeV.
The data correspond to an integrated luminosity of 35.9\fbinv 
collected with the CMS detector at the LHC.

The electroweak production of two \PZ bosons in association with two jets
was measured with an observed (expected) significance of 2.7 (1.6) standard
deviations. The fiducial cross section is
${\sigma_{\textrm{EW}}(\Pp\Pp\rightarrow\textrm{ZZjj}\rightarrow\ell\ell\ell^{\prime}\!\ell^{\prime}\textrm{jj}) = 0.40^{+0.21}_{-0.16}\stat~^{+0.13}_{-0.09}\syst\unit{fb}}$, consistent with the standard model prediction of ${0.29^{+0.02}_{-0.03}\unit{fb}}$.

Limits on anomalous quartic gauge couplings were set at the 95\% confidence level in terms of effective field theory
operators, in units of TeV$^{-4}$: 
\begin{center}
${-}0.46 < f_\textrm{T0}/\Lambda^4 <0.44$\\
${-}0.61 < f_\textrm{T1}/\Lambda^4 <0.61$\\
$\x{-}1.2 < f_\textrm{T2}/\Lambda^4 < 1.2\x$ \\
${-}0.84 < f_\textrm{T8}/\Lambda^4 <0.84$\\
\y$\x{-}1.8  < f_\textrm{T9}/\Lambda^4 <1.8\x$.
\end{center}

These are the first results for the electroweak production of two $\PZ$ bosons in association with jets at the LHC and the most stringent limits on the T0, T1, T2, T8, and T9 anomalous quartic gauge couplings to date.

\begin{acknowledgments}

We congratulate our colleagues in the CERN accelerator departments for the excellent performance of the LHC and thank the technical and administrative staffs at CERN and at other CMS institutes for their contributions to the success of the CMS effort. In addition, we gratefully acknowledge the computing centres and personnel of the Worldwide LHC Computing Grid for delivering so effectively the computing infrastructure essential to our analyses. Finally, we acknowledge the enduring support for the construction and operation of the LHC and the CMS detector provided by the following funding agencies: BMWFW and FWF (Austria); FNRS and FWO (Belgium); CNPq, CAPES, FAPERJ, and FAPESP (Brazil); MES (Bulgaria); CERN; CAS, MoST, and NSFC (China); COLCIENCIAS (Colombia); MSES and CSF (Croatia); RPF (Cyprus); SENESCYT (Ecuador); MoER, ERC IUT, and ERDF (Estonia); Academy of Finland, MEC, and HIP (Finland); CEA and CNRS/IN2P3 (France); BMBF, DFG, and HGF (Germany); GSRT (Greece); OTKA and NIH (Hungary); DAE and DST (India); IPM (Iran); SFI (Ireland); INFN (Italy); MSIP and NRF (Republic of Korea); LAS (Lithuania); MOE and UM (Malaysia); BUAP, CINVESTAV, CONACYT, LNS, SEP, and UASLP-FAI (Mexico); MBIE (New Zealand); PAEC (Pakistan); MSHE and NSC (Poland); FCT (Portugal); JINR (Dubna); MON, RosAtom, RAS, RFBR and RAEP (Russia); MESTD (Serbia); SEIDI, CPAN, PCTI and FEDER (Spain); Swiss Funding Agencies (Switzerland); MST (Taipei); ThEPCenter, IPST, STAR, and NSTDA (Thailand); TUBITAK and TAEK (Turkey); NASU and SFFR (Ukraine); STFC (United Kingdom); DOE and NSF (USA).

\hyphenation{Rachada-pisek} Individuals have received support from the Marie-Curie programme and the European Research Council and Horizon 2020 Grant, contract No. 675440 (European Union); the Leventis Foundation; the A. P. Sloan Foundation; the Alexander von Humboldt Foundation; the Belgian Federal Science Policy Office; the Fonds pour la Formation \`a la Recherche dans l'Industrie et dans l'Agriculture (FRIA-Belgium); the Agentschap voor Innovatie door Wetenschap en Technologie (IWT-Belgium); the Ministry of Education, Youth and Sports (MEYS) of the Czech Republic; the Council of Science and Industrial Research, India; the HOMING PLUS programme of the Foundation for Polish Science, cofinanced from European Union, Regional Development Fund, the Mobility Plus programme of the Ministry of Science and Higher Education, the National Science Center (Poland), contracts Harmonia 2014/14/M/ST2/00428, Opus 2014/13/B/ST2/02543, 2014/15/B/ST2/03998, and 2015/19/B/ST2/02861, Sonata-bis 2012/07/E/ST2/01406; the National Priorities Research Program by Qatar National Research Fund; the Programa Clar\'in-COFUND del Principado de Asturias; the Thalis and Aristeia programmes cofinanced by EU-ESF and the Greek NSRF; the Rachadapisek Sompot Fund for Postdoctoral Fellowship, Chulalongkorn University and the Chulalongkorn Academic into Its 2nd Century Project Advancement Project (Thailand); and the Welch Foundation, contract C-1845.

\end{acknowledgments}

\bibliography{auto_generated} 
\cleardoublepage \appendix\section{The CMS Collaboration \label{app:collab}}\begin{sloppypar}\hyphenpenalty=5000\widowpenalty=500\clubpenalty=5000\textbf{Yerevan Physics Institute,  Yerevan,  Armenia}\\*[0pt]
A.M.~Sirunyan, A.~Tumasyan
\vskip\cmsinstskip
\textbf{Institut f\"{u}r Hochenergiephysik,  Wien,  Austria}\\*[0pt]
W.~Adam, F.~Ambrogi, E.~Asilar, T.~Bergauer, J.~Brandstetter, E.~Brondolin, M.~Dragicevic, J.~Er\"{o}, M.~Flechl, M.~Friedl, R.~Fr\"{u}hwirth\cmsAuthorMark{1}, V.M.~Ghete, J.~Grossmann, J.~Hrubec, M.~Jeitler\cmsAuthorMark{1}, A.~K\"{o}nig, N.~Krammer, I.~Kr\"{a}tschmer, D.~Liko, T.~Madlener, I.~Mikulec, E.~Pree, D.~Rabady, N.~Rad, H.~Rohringer, J.~Schieck\cmsAuthorMark{1}, R.~Sch\"{o}fbeck, M.~Spanring, D.~Spitzbart, W.~Waltenberger, J.~Wittmann, C.-E.~Wulz\cmsAuthorMark{1}, M.~Zarucki
% ...
\vskip\cmsinstskip
% ...
1:~~Also at Vienna University of Technology, Vienna, Austria\\
% ...

\end{sloppypar}
\end{document}